\documentclass{iopart}
\input epsf

\newcommand{\siml} {\,{}^<_{\sim}\,}
\newcommand{\simg} {\,{}^>_{\sim}\,}

\begin{document}
\jl{6}

\title{From (p)reheating to nucleosynthesis}
\author{Karsten Jedamzik$^{1,2}$}

\address{$^1$ Physique Math\'ematique et Th\'eorique, 
Universit\'e Montpellier II, 
34095 Montpellier, France}
\vskip 0.2in
\address{$^2$ Max-Planck-Institut f\"ur Astrophysik, 85741 Garching, Germany}

\date{\today}

\begin{abstract}
This article gives a brief qualitative description of the possible
evolution of the early Universe between the end of an inflationary epoch
and the end of Big Bang nucleosynthesis. After a general introduction,
establishing the minimum requirements cosmologists impose on this cosmic
evolutionary phase, namely, successful baryogenesis, the production
of cosmic dark matter, and successful light-element nucleosynthesis,
a more detailed discussion on some recent developments follows. 
This latter includes the physics of preheating, the putative 
production of (alternative) dark matter, and  
the current status of Big Bang nucleosynthesis. 
\footnote{Article based on a 
talk presented at ``The Early Universe and Cosmological Observations: a 
Critical Review'', Cape Town, July 2001}
\end{abstract}

\section{Introduction}
I was asked by the organizers of the conference to present a review 
of processes occurring in the early Universe, between the end of a 
cosmological inflationary period and the end of Big Bang nucleosynthesis.
Due to the multitude of possible processes and phenomena which may
have occurred over such a long evolutionary interval, and the plethora
of scientific work which exists on the subject, accomplishing such a
task on a few pages is a difficult undertaking and bound to be done 
superficially.
Nevertheless, I realize that such a mini-review of reviews (i.e. frequent
citing of other review articles) may be of some aide to scientists outside
of the community and provide an initial starting point for further study.
It is in this way I like the article to be understood. My apology goes to
all scientists whose work has not been mentioned explicitly.

The article is split into two parts. The first part (sections 2 and 3)
contains a very general account of the evolution of the early Universe
and the processes which must have occurred to create an ``acceptable''
Universe. In the second part (sections 4 - 9)
more emphasis is laid on some recent developments and advances in
the field. Note that the article does not consider cosmological
phenomena due to the (putative) existence of 
extra spatial dimensions (cf. David Wands, these proceedings), 
as the study of such is still in it's infancy.

\begin{figure}[t]
\centering 
\leavevmode\epsfysize=12cm \epsfbox{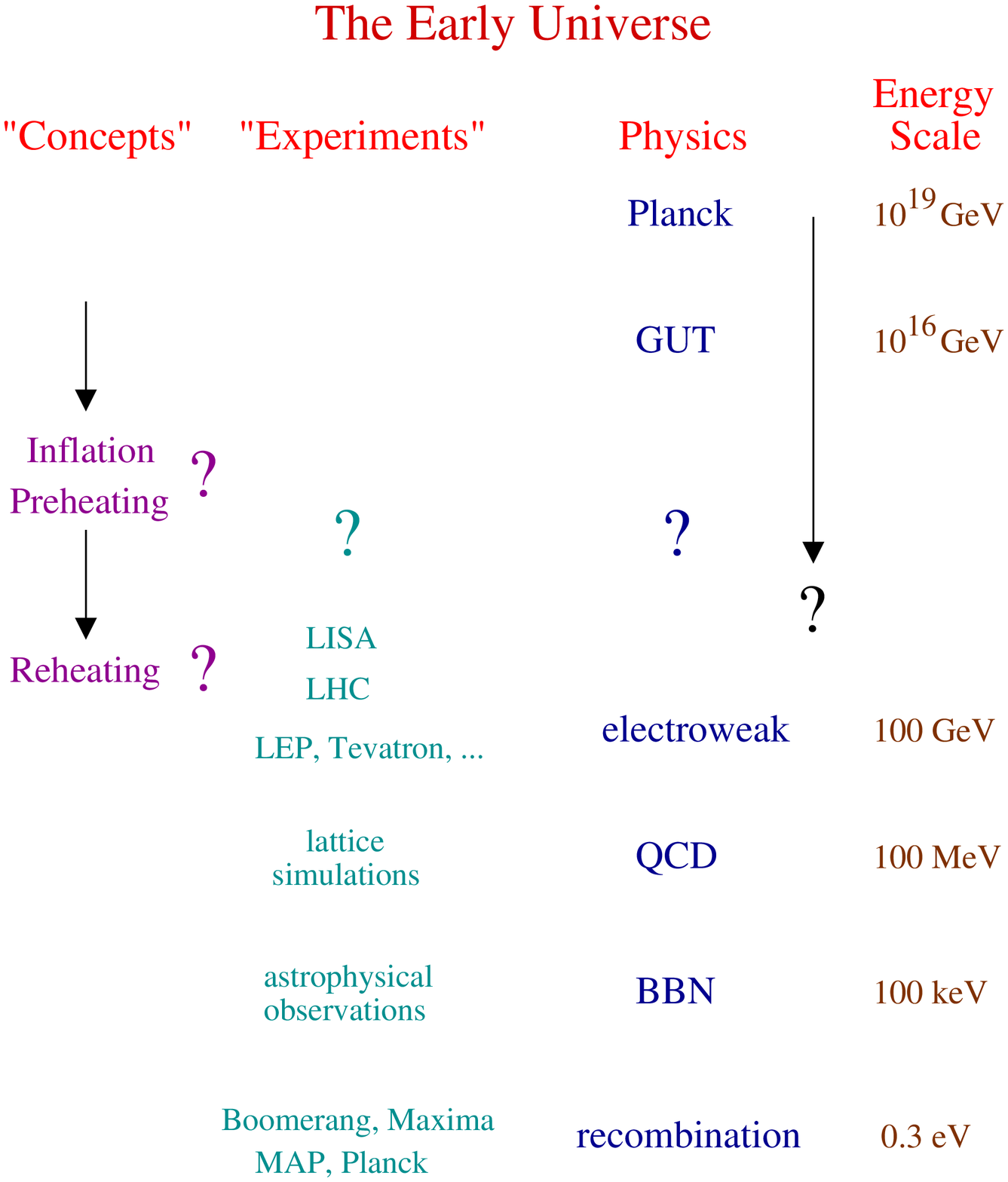}\\ 
\end{figure} 

\section{The Evolution of the Early Universe: a General Overview}

The hot Big Bang model of the Universe is subject to overwhelmingly 
strong support by cosmologists. This is mainly due to three cosmological
observations: the recession of galaxies around the Milky Way according
to Hubble's law, the existence of the cosmic microwave background radiation 
(hereafter; CMBR)
with an almost flawless blackbody spectrum at temperature $T \approx 2.726\,$K,
and the successes of explaining observationally inferred light-element
abundances in old, ``uncontaminated'' astrophysical environments 
by cosmological production during a Big Bang nucleosynthesis 
(hereafter; BBN) epoch occurring at 
$T\approx 100\,$keV. Additional support to the hot Big Bang is lend due to
our increasingly accurate understanding of how cosmic structure formed from
initially small perturbations extant at recombination at $T\approx 0.3\,$eV,
as witnessed by fluctuations in the CMBR. When considering the hot 
Big Bang model in the sixties it was immediately realized that in 
it's very early stages the Universe may have been extremely hot. 
The above table provides
a summary of the possible physics, and the respective energy scale, 
the early Universe may have been subject to in its early stages 
(cf.~\cite{KT90,B90}).
Also shown is the phenomenologically successful concept of inflation 
(cf. Andrew Liddle, these proceedings), which may be followed by an epoch of 
preheating, and requires an epoch of reheating, in order to produce 
an ``acceptable'' (our) Universe. 
It is termed ``concept'' rather than
``physics'' as it is currently not known, 
assuming inflation took place, at which
energy scale it occurred, and in the context of which 
particle physics model. 
Thus, of course, there is considerable room for speculation about the evolution
of the early Universe, in particular also, since particle physics is known
in detail only up to the electroweak scale.

Nevertheless, note the breathtaking variety of ``experiments'' and 
methods which are, and will be, used to gain insight about the early Universe.
These include observations of CMBR anisotropies by 
balloon-borne experiments,
such as Boomerang and Maxima, and upcoming satellite missions, i.e. MAP and
Planck (cf. David Langlois and Anthony Lasenby, these proceedings),
observations of absorption and emission lines in gaseous nebulae,
such as clouds along the line of sight towards a quasar (quasar absorption
line systems) and compact blue galaxies (cf. Sec. 9), lattice gauge
simulations to find out about the order of the QCD phase transition
(which is still unknown!), existing (LEP, Tevatron) and planned (LHC)
particle accelerators which have the potential to discover the ubiquitous
Higgs particle and/or supersymmetry, as well as space-born gravitational
wave interferometry (LISA), 
which could detect gravitational waves right
form the very early Universe. 

\section{Cosmologist's Minimum Requirements on the Early Universe
after Inflation}

An early evolutionary stage of the Universe different from the present
has the benefit of providing explanation of why exactly 
the Universe is as observed today. In particular, properties of the 
Universe, which theretofore
could only be attributed to initial conditions, may be explained by
physical processes which occurred during the hot Big Bang.
The virtue of an early cosmic inflationary phase to free the Universe
from ``unwanted'' relics (e.g. magnetic monopoles produced during a GUT 
era) as well as explaining present cosmic large-scale homogeneity
and apparent close-to-flatness has been well established (cf.~\cite{KT90}).
After this process cosmic energy density of the Universe is dominated
by oscillations of the inflaton condensate (or other bosonic fields coupled
to the inflaton). These condensates have to
(essentially) completely decay into particles which couple to the standard 
model particles, in order for energy density to be dominated by ordinary
radiation first (which ultimately becomes the CMBR), 
and dark matter later (at $T\siml 1\,$eV). Moreover, it is 
believed that the radiation has to be well equilibrated by the onset of BBN
($T\sim 1\,$MeV). On the other hand, there are good arguments, 
assuming the existence of supergravity and gravitinos, that the Universe
should have not reached temperatures as high as $T\sim 10^8 - 10^9\,$GeV
(cf. Sec. 6). 
The process of inflaton decay may occur via individual inflaton-particle 
decay (reheating), and may (or may not) be preceded by a preheating era.
During preheating other bosonic (or fermionic) degrees of 
freedom coupled to the
inflaton are efficiently excited via parametric resonance due to the 
coherence of the oscillating inflaton condensate (cf. Sec. 5). 

Irrespective of the thermal history of the Universe between the end of
inflation and the end of the BBN freeze-out, there exist three 
minimum requirements on this evolutionary phase. An asymmetry between 
matter and antimatter has to be dynamically generated (baryogenesis),
the production of a component which becomes to dominate the energy density
at late times and accounts for the right cosmic structure formation 
(dark matter) has to occur, and the cosmic conditions during BBN
have to be such that light-element synthesis occurs in accord with
observationally inferred light-element abundances. All three may (or may not) 
be accomplished by various possibilities and scenarios of which some
are shown on the following page. 

For recent reviews on baryogenesis the reader
is referred to Ref.~\cite{baryo}, and on dark matter to Ref.~\cite{DM}
(mostly supersymmetric dark matter).
References to reviews on BBN, standard and non-standard, 
are given in Sec. 9. Finally, magnetic fields
(cf.~\cite{magnetic} for review) and gravitational waves
(cf.~\cite{M00} for review) may also play a significant role in early 
Universe cosmology.

\begin{figure}[t]
\centering 
\leavevmode\epsfysize=10cm \epsfbox{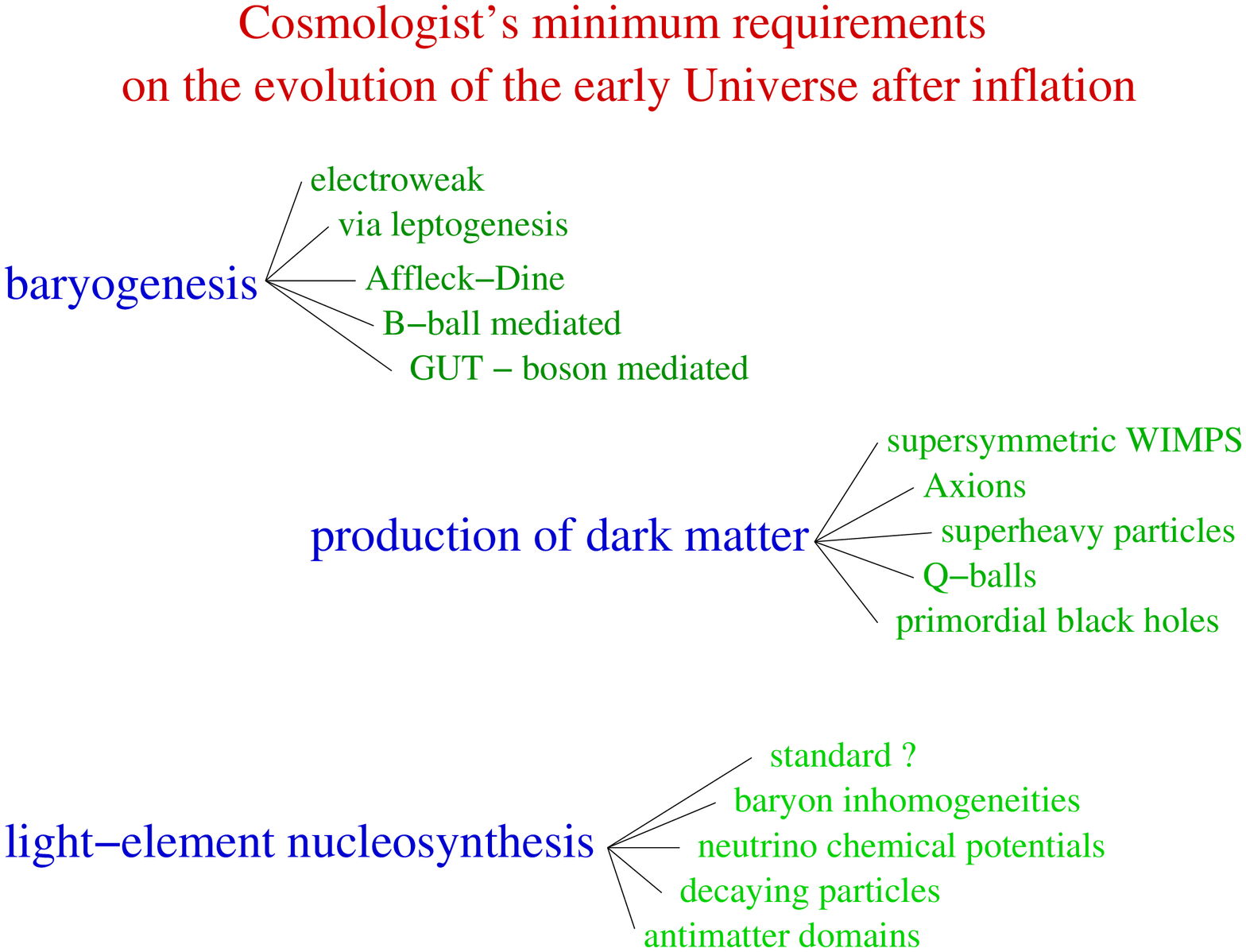}\\ 
\end{figure}

\section{Cosmological effects due to violation of adiabaticity}

Observable remnants of early cosmic evolution are generally due to
deviations from equilibrium.
A number of such processes, having been recently proposed in the literature, 
may be attributed to non-trivial ``particle'' behavior
due to violation of adiabaticity of mode frequencies.
Consider a non-interacting, massive scalar field in the expanding 
Universe~\cite{BD84}.
The usual expansion of the field operator $\hat\chi$ in Fourier space
\begin{eqnarray}
\hat \chi(t, {\bf x}) =
{1\over{(2\pi)^{3/2}}} \int d^3k\ {\Bigl( \hat a_{{k}} \chi_{
k}(t)\, e^{ -
i{{\bf k}}{{\bf x}}}
+ \hat a_{{k}}^ + \chi^*_k(t)\, e^{i{k}{{\bf x}}}
\Bigr)}\, ,
\end{eqnarray}
leads to the equation of motion for the amplitudes
\begin{eqnarray}
\ddot \chi_k + 3{{\dot a}\over a}\dot \chi_k + {\left(
{{\bf k^2}\over a^2}
 + m_{\chi}^2\right)} \chi_k = 0 \, ,
\label{EOM}
\end{eqnarray}
where $a$ is cosmic scale factor, $k/a$ and $m_{\chi}$
are physical momentum and mass of the particle, respectively, and
a dot indicates a derivative with respect to cosmic time $t$.
Solutions to Eq.~(\ref{EOM}) may be found in a WKB perturbation scheme 
($\dot{\omega}/\omega^2\ll 1$) and are given by
\begin{eqnarray}
\chi_k(t) = \tilde{\chi}_k(t)\,
{\rm exp}^{\pm i\int dt\,\omega_k} \, ,
\end{eqnarray}
with $\omega_k =\sqrt{(k/a)^2 + m_{\chi}^2}$ particle mode frequencies,
and $\tilde{\chi}_k(t)$ slowly varying functions. These functions
are given by $\chi_k^0/a^{3/2}$ for nonrelativistic 
particles ($k/a \ll m_{\chi}$, NR) 
and $\chi_k^0/a$ for ultrarelativistic particles
($k/a \gg m_{\chi}$, UR), where $\chi_k^0$ is constant. 
The energy density per mode is given by
\begin{eqnarray}
\epsilon_k = {1\over 2}|\dot{\chi}_k|^2 + {1\over 2}\omega_k^2|{\chi}_k|^2
= \omega_k n_k^0/a^3\, ,
\end{eqnarray}
where $n_k^0$ may be interpreted as conserved particle 
occupation number 
given by $n_k^0 = m_{\chi}|\chi_k^0|^2$ (NR) and
$n_k^0 = k|\chi_k^0|^2$ (UR). The total energy
density may be computed by
\begin{eqnarray}
\epsilon = {1\over (2\pi)^3}\int d^3k\, \omega_k^2\, |{\chi}_k|^2\nonumber 
\end{eqnarray}
and exhibits the well-known redshifting by $\sim 1/a^4$ for 
relativistic matter, 
and by $\sim 1/a^3$ for nonrelativistic matter, 
due to the expansion of the Universe.
However, this intuitive picture of particle behavior
breaks down whenever the WKB condition ${\dot{\omega}_k/ \omega_k^2} \ll 1$ 
is violated. In this case, particle occupation number is not conserved.
This may occur, for example, during

\vskip 0.1in
\noindent
{\bf preheating:} Given the potential
\begin{equation}
V = {1\over 2} (m_{\phi}^2\phi^2 + g^2\phi^2\chi^2 +
m_{\chi}^2\chi^2)\, ,
\label{potential} 
\end{equation}
where $\phi$ is the inflaton field and $\chi$ a 
``matter''field coupled to it, a $\chi$ effective mass 
$m_{\chi ,eff}^2 = m_{\chi}^2 + g^2\phi^2$ arises. At the end of inflation,
when $\phi$ oscillates around the minimum of the potential
(i.e. $\phi (t) = \Phi (t)\,{\rm sin}( m_{\phi}\, t)$)
an effective, and time-dependent contribution 
to the squared $\chi$ mass of 
$g^2\phi^2 = g^2\Phi^2(t)\,{\rm sin}^2( m_{\phi}\, t)$ 
results~\cite{preh,field}.
Depending on the coupling constant $g$ this effective mass may be
appreciable since, for example at the end of chaotic inflation,  
$\Phi (t) \approx 0.3M_{pl}$ (and $m_{\phi} \approx 10^{-6}M_{pl}$),
where $M_{pl}$ denotes the Planck mass.
The maximum violation of adiabaticity obtains when $\phi$ passes through 
zeros: ${\dot{\omega}_k/ \omega_k^2}\big|_{\rm max} 
\approx q^{1/2}m_{\phi}^2/((k/a)^2 + m_{\chi}^2)$ which may be large,
particularly when the resonance parameter 
$q = g^2\Phi^2(t)/4m_{\phi}^2$ is large.
In this case parametric resonance phenomena occur which may be
associated with efficient explosive $\chi$ particle production due
to nonperturbative quantum effects, driven by the coherent oscillations
of the inflaton condensate.  

\vskip 0.1in
\noindent
{\bf the rapid cosmic expansion during and at the end of inflation:}
The Hubble constant at the end of inflation may be extremely large, e.g.
$H = (\dot{a}/a)\approx 3 m_{\phi}(\Phi /M_{pl}) \sim 10^{13}{\rm GeV}$ 
for chaotic inflation. This leads to violation of the WKB condition
${\dot{\omega}_k/ \omega_k^2}\, {}_{\sim}^>\, 1$ for all modes 
with $((k/a)^2 + m^2_{\chi}) 
\, {}_{\sim}^<\, H^2$, i.e. even for very massive particles.
Such particles may thus be created out of the vacuum 
due to the rapid expansion of the Universe~\cite{CKR99,KT98}, even when no
explicit coupling to the inflaton exists.

\vskip 0.1in
\noindent
In the Sec. 5 and 6 some possible cosmological implications of these
processes are summarized.

\section{Preheating}
\begin{figure}[t]
\centering 
\leavevmode\epsfysize=7cm \epsfbox{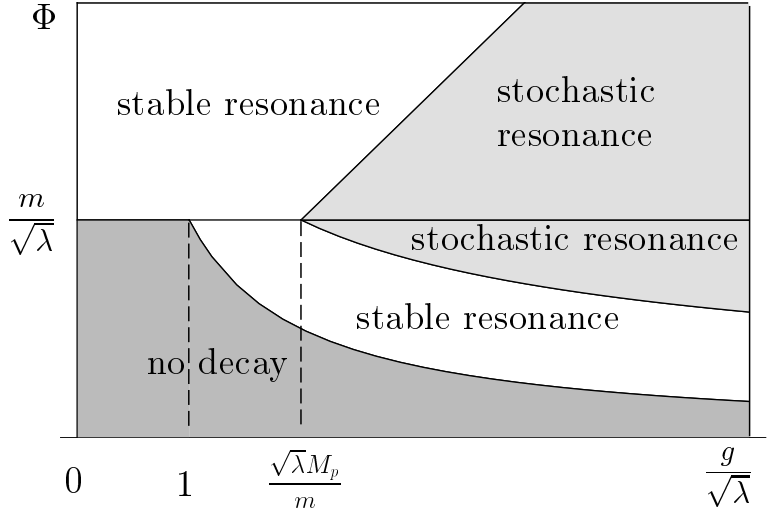}\\ 
\caption[scdm]{The nature of preheating instabilities in different parts
of the parameter space for a theory with potential
$V = (m^2/2)\phi^2 +(\lambda/4)\phi^4 + (g^2/2)\phi^2\chi^2$. The reader
is referred to Ref.~\protect\cite{GKLS97} for detailed explanation.}
\label{fig3}
\end{figure}

Here I discuss the physics of preheating in a model with
potential as in Eq.~(\ref{potential}), and in the limit of small 
$m_{\chi}$. This model is known to exhibit stochastic 
resonance~\cite{stochastic}. 
Though preheating may occur in other models via physically
different instabilities,
e.g. ``stable'' preheating in ``conformally invariant'' models with potential 
$V = (\lambda/4)\phi^4 + (g^2/2)\phi^2\chi^2$~\cite{GKLS97}, 
or ``tachyonic''
preheating in hybrid inflationary models~\cite{tachyonic},
or other mechanism~\cite{other}, 
the basic conclusions 
and implications often remain the same. 
The nature of the instabilities
for different parameter combinations of a particular model are shown
in Fig.~\ref{fig3}.

\subsection{preheating: stages I, II, and III}

\vskip 0.1in
\noindent
Stage I: Right after the termination of inflation,
the number of bosonic ``matter'' $\chi$ quanta coupled to the
inflaton grows exponentially via parametric resonance
\begin{equation} 
\chi_k \propto {\rm exp}(\mu_k m_{\phi}\, t/2)\, ,
\end{equation} 
driven by the coherent oscillations of the inflaton condensate. 
This may happen even for a $\chi$ field initially in its
vacuum state. In the above expression $\mu_k$ is a numerical constant 
(Floquet index) which depends
on wave vector $k$. Usually, efficient conversion of inflaton energy into
$\chi$-quanta obtains only in the
broad resonance regime with resonance parameter 
$q = g^2\Phi(t)^2/4m_{\phi}^2 \approx (g/10^{-3})^2 10^4 > 1$
(cf. Sec. 4 for notation), where the second equation assumes chaotic inflation. 
In this case, there exist many, broad bands in k-space of
sufficiently unstable ($\mu_k > 0$) modes. Due to phase space arguments
most of the inflaton energy is ``dumped'' into modes with wavevectors
around the largest, still unstable, 
wavevector, i.e. $k_{\rm max}\approx q^{1/4}m_{\phi}$.
The initial decay of the energy density in the inflaton condensate 
due to preheating is often much
more efficient than that by perturbative inflaton decay (i.e. reheating). 
The decay rate of the latter process is given by 
$\Gamma (\phi\to\psi\bar{\psi}) = h^2m_{\phi}/8\pi$ 
for decay of the inflaton into a fermion ($\psi\bar{\psi}$)
pair via the interaction $L_{int}\sim -h\phi\bar{\psi}\psi$.
Production of fermions via preheating has also been 
considered~\cite{fermions}, it is,
nevertheless, comparatively inefficient as fermion production is subject
to constraints from the Pauli exclusion principle, whereas similar does not
hold for bosons.

\vskip 0.1in
\noindent
Stage II: After only a number of inflaton oscillations 
the energy density in the produced  $\chi$ quanta may become considerable.
A second stage of preheating commences when the existence of $\chi$ quanta
``backreacts'' on the dynamics of the inflaton 
background~\cite{stochastic}, inducing
an effective inflaton mass $g^2<\chi^2> \simg m_{\phi}^2$ by virtue of the
interaction term in the potential Eq.~(\ref{potential}).
This increases the inflaton oscillation frequency and decreases the value
of the resonance parameter $q$.
After only a small number of oscillations $<\chi^2> \approx
\Phi^2$, the resonance parameter becomes smaller than unity 
(i.e. $q\sim 1/4$), and parametric resonance becomes inefficient.
The existence of $\chi$ quanta also influences
the inflaton background by $\chi$ quanta ``rescattering'' on the inflaton 
background, thereby creating $\phi$ quanta and destroying the coherence
of the inflaton condensate.

\vskip 0.1in
\noindent 
Stage III: After the end of the second stage, resonance has become
inefficient in redistributing energy between the inflaton and the
``matter'' field. At this point, the Universe is described 
by a highly nonthermal distributions of bosons (with energy
mostly in infrared modes).  
These distributions thermalize subsequently partially 
via turbulent wave dynamics, as recently shown by direct numerical
simulations~\cite{Npreheat}. Nevertheless, in most models reheating, i.e. 
existence of a well-equilibrated particle distribution as well as the
complete decay of the inflaton energy, is still not complete,
and may often only be accomplished by the decay of 
bosons to fermions (as in the standard theory of reheating).

\subsection{metric preheating}
Perturbations $\varphi$ of the metric around the Friedann-Robertson-Walker 
metric are sourced by perturbations $\delta\phi_{I}$ in the matter fields
\begin{equation}
\dot{\varphi}_k + H\varphi_k = {4\pi\over M_{pl}^2}\sum_I 
\dot\phi_I\delta\phi_{Ik}\, ,
\label{metric}
\end{equation}
as shown here via the Einstein constraint equation in longitudinal gauge.
Thus, parametric amplification (exponential growth) could generically also 
occur for metric fluctuations when parametric amplification occurs in the
``matter'' sector. 
Moreover, it has been stressed that resonance often occurs 
for the $k\to 0$ mode, which may correspond 
to cosmologically relevant scales (e.g. the observed scales of the
CMBR)~\cite{metric}. 
It is well known, that inflation may offer
an attractive explanation for the density (metric) fluctuations on
cosmological scales. This raises the question, if 
an epoch of preheating may actually destroy this prediction for the 
metric perturbations produced during inflation ? The answer to this
question is, in principle yes, though in practice it seems to 
occur only within a small subset of models~\cite{metric1}. This may happen, 
in particular, in the few cases when a potentially existing background 
field $\phi_I$ of the parametrically amplified $\delta\phi_{Ik}$ is not
damped during inflation (cf. Eq.~(\ref{metric})
\footnote{Note that in the opposite case ($\phi_I=0$)
metric fluctuation may only be generated to second order in 
$\delta\phi_{Ik}$}), but may {\it not} when
$\phi_I$ is the inflaton itself~\cite{AB01}(i.e. single-field preheating).  
A second question is if \lq\lq small-scale\rq\rq\ metric fluctuations 
produced during preheating
may induce the efficient formation of primordial black holes ?
Finding an answer to this question is technically involved as 
it requires full numerical simulations of the Einstein
equations~\cite{prePBH1}, and, again, model-dependent.
Thus, contrary conclusions (attained by varying degrees of
sophistication) have been reached in the literature~\cite{prePBH2}.

\subsection{cosmological implications of preheating}

Though the Universe is, of course, not required to pass through
an epoch of preheating, if it occurrs it may have several observable
consequences. One obvious observation is that the Universe
may have been left for a very long time in a very non-equilibrium state 
between the end of inflation and the onset of an ordinary (thermalized) 
radiation dominated epoch. During this non-equilibrium phase the generation
of dark matter but also baryogenesis
may be accomplished even when radiation domination commences at temperatures
much below the electroweak scale. Baryogenesis may proceed via excitation
of non-thermal (baryon-number violating) sphaleron transitions induced
by the large number of produced infrared quanta~\cite{prebaryo}. 
If the Universe then reheats to a temperature well below the 
electroweak scale the common problem 
(in ordinary electroweak baryogenesis) of subsequent wash-out of the generated
baryon number right below the transition may be circumvented.
Baryogenesis may also occur in a different scenario, via the production of
massive $\sim 10^{15}\,$GeV GUT bosons (see below), 
and the subsequent inevitable baryon-number violating decay~\cite{GUTbaryo}.
Old attempts to employ such scenarios were problematic due to
increasingly large cosmic temperatures required for production of GUT bosons,
and the possible concomitant overproduction of gravitinos and/or 
magnetic monopoles. With the aid of non-equilibrium processes during 
preheating particles of mass $M$ may be efficiently generated, even if
cosmic energy density $\epsilon$ after inflation 
($\sim (10^{13}\,{\rm GeV})^4$ in chaotic inflation)
falls well below $M^4$.
This is possible since parametrically resonant production of particles
may occur whenever $M\siml \sqrt{g\, m_{\phi}\, \Phi} \sim 10^{15}{\rm
GeV}$ (cf. Sec. 4 - preheating).
Moreover, even relics of mass close to the Planck scale may be generated
when quanta with effective mass 
$\sim g\Phi \sim 10^{17} - 10^{18}{\rm GeV}$ (due to the interaction term 
in Eq.~\ref{potential}) decay immediately after
production~\cite{heavy}. 
If such particles are unstable, but long-lived, they may
in fact be the primaries of the observed highest energy
cosmic rays~\cite{HECR}. Furthermore, the exceedingly large fluctuations in
bosonic fields $\chi$ coupled (parametrically unstable) to the inflaton, i.e.  
$\sqrt{<\chi^2>}\sim 10^{16}{\rm GeV}$ may induce non-thermal phase 
transitions which, in turn, may result in production of domain walls, 
strings, and monopoles (e.g. at the GUT scale)~\cite{phase}. 
Last but not least, as discussed above (Sec. 5.2) 
production of primordial black holes and/or 
generation of metric fluctuations on cosmologically relevant scales may
also result. Many of these implications may be used to constrain the 
coupling of the inflaton to other matter fields.

\section{Particle production in the early Universe}

\subsection{the moduli \& gravitino problems}

Moduli are light scalar fields with weak 
coupling to other particles. As will be seen, they are not easily
diluted during inflation (in contrast to massive degrees of freedom).
This poses a problem, as they may become to dominate cosmic energy
density early on~\cite{moduli}. 
Furthermore, if their masses are of the weak scale
and they have Planck-suppressed couplings to other particles, they
may decay after BBN, thereby disrupting light-element yields in
an unacceptable way (cf. Sec. 6.1.3.). 
Supersymmetric particle theory generically contains 
flat directions in the scalar potential, which may be associated 
with moduli fields of exactly those properties.

\subsubsection{classical moduli problem:} Consider the equation of motion for
a classical massive, and non-interacting field
\begin{eqnarray}
\ddot \chi + 3H\dot \chi 
 + m^2_{\chi}\chi = 0\, .
\label{classical}
\end{eqnarray}
Eq.~(\ref{classical})
allows for (undamped) solutions $\chi\approx {\rm constant}$
when $m_{\chi} \ll H$ is satisfied. As the Hubble constant 
may be significant during inflation, i.e. $H\sim 10^{13}\,$GeV in the
simplest single-field scenarios, even fields with comparatively large
masses are not damped during inflation. This defeats one
of the original successes of inflationary theory, in particular,
the creation of a ``clean'' Universe by dilution
of initially present unwanted relics into oblivion. 
One may pose that $\chi = 0$ initially, however, the problem
reappears also at the quantum level.

\subsubsection{quantum moduli problem:} Moduli quanta with 
$\sqrt{(k/a)^2 + m^2_{\chi}} < H^2$ 
are continuously produced during, and at the end of inflation~\cite{qmoduli}.
Even if the initial state at the beginning of inflation was,
for some unknown reason, in the $\chi$ vacuum state, quanta are
reproduced due to the rapid expansion of the Universe.
This process is not unlike the generation of density perturbations
from vacuum fluctuations in the inflaton field, 
believed to be responsible for cosmological
structure formation at some later epoch.
It may be shown~\cite{qmoduli1}
that the fractional contribution of generated $\chi$ quanta to the
present critical density is appreciable
\begin{equation} \!\!\!\!\!\!\!\!\!\!\!\!
\Omega_{\chi}^0 \sim \Omega_{\gamma}^0\bigl(H_I/M_{pl}\bigr)^2
\bigl(T_{\rm RH}/T_0\bigr) 
\sim 10^7\bigl(H_I/10^{13}{\rm GeV}\bigr)^2
\bigl(T_{\rm RH}/10^{9}{\rm GeV}\bigr)\, ,
\label{Omega}
\end{equation}
unless the Hubble constant during inflation $H_I$ is comparatively
small, and/or the reheat temperature $T_{\rm RH}$ of the Universe
is low. In the above, $\Omega_{\gamma}^0$ and $T_0$ denote 
present
contribution of radiation energy density to the critical density and
present CMBR temperature, respectively. Note that Eq.~(\ref{Omega})
assumes redshifting of cosmic inflaton
energy density after inflation, and before beginning of 
radiation domination at $T_{\rm RH}$ as $\rho_I \sim a^{-3}$
(e.g. for inflation with potential $m_{\phi}^2\phi^2/2$).
\footnote{For $\rho_I \sim a^{-4}$ resulting from potentials $\lambda 
\phi^4/4$, the reheat temperature $T_{\rm RH}$ in Eq.~(\ref{Omega}) may
be roughly substituted by the value of $\lambda^{1/4}\phi$ at the end of
infation.}  
The problem exemplified by Eq.~(\ref{Omega}) may be somewhat alleviated
when either bosonic fields are conformally coupled to gravity
(i.e. extra term of $(1/12)R\chi^2$ in the Lagrangian, where $R$ is the
Ricci scalar), or when fermions are considered. 
In this case, Eq.~(\ref{Omega}) should be multiplied by a factor 
$\sim 10^{-5}m_{\chi}/m_{\phi}$.

\subsubsection{gravitino problem:} 
\begin{figure}[t]
\centering 
\leavevmode\epsfysize=9cm \epsfbox{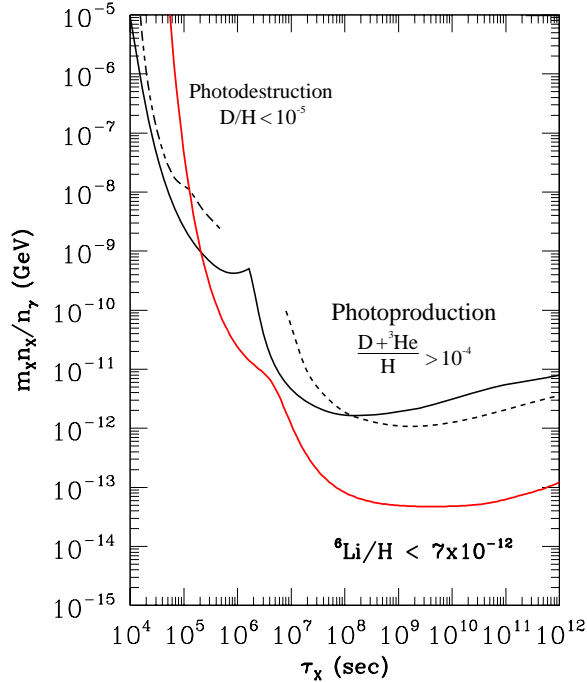}\\ 
\caption[scdm]{Limits on the abundance of a relic, decaying particle with
mass $m_X$ and density $n_X$, relative to cosmic photon density 
$n_{\gamma}$, as a function of decay time. Limits based on D
and $^3$He isotope production/destruction (upper lines)
are taken from Ref.~\protect\cite{S96},
whereas possible limits based on $^6$Li production (lower lines)
are taken from Ref.~\protect\cite{J00}.}
\label{fig4}
\end{figure}

Gravitinos are spin (3/2) particles
predicted to exist in phenomenologically successful
locally supersymmetric (supergravity) theories. Their (possible)
existence implies constraints on early Universe evolution~\cite{gravitinos}.
At low energy, gravitino interactions are weak, suppressed by powers 
of $M_{pl}$. In gravity-mediated supersymmetry breaking gravitino masses of
$M_G\, {}^<_{\sim}\, 1\,$TeV are strongly favored. Unless they are
the lightest supersymmetric particle (LSP), they decay into
standard model particles and their supersymmetric partners
(i.e. $G\to \gamma\bar{\gamma}$ and $G\to g\bar{g}$ where $\gamma$
is a photon and $g$ is a gluon). This typically happens after the
epoch of BBN, with half life 
$\tau_G \approx M_{pl}^2/M_G^3\sim 10^{5}{\rm sec}\, 
(M_G/{\rm 1\,TeV})^{-3}$. Such decay results in destruction
of the BBN synthesized deuterium, overproduction of
$^3$He, or even more drastic, 
overproduction of $^6$Li 
These constraints are summarized 
in Fig.~\ref{fig4}.
Since gravitinos may be recreated 
after inflation by collisions of particles in a hot plasma, one may impose
an upper limit on the reheat temperature of the Universe. This upper limit
is approximately $T_{\rm RH} \siml 10^8 - 10^9\,$GeV 
(\cite{S96} and references therein), imposing stringent
constraints on the thermal history of the early Universe. If instead
$M_G\sim 100\,$keV and the gravitino is the LSP 
(as typical in gauge-mediated supersymmetry breaking) a limit of
$T_{\rm RH} \siml 10^4\,$GeV may be imposed from their maximum possible
contribution to the present density.

Nevertheless, it has been recently realized that gravitino production even
{\it before} radiation domination via non-thermal processes may be very 
efficient, possibly ruling out
whole classes of early Universe scenarios. Gravitinos may be created
out of vacuum fluctuations due to the rapid expansion of the
Universe. More importantly, 
the spin (1/2) component of gravitinos (which is associated with the 
goldstino) could be produced immediately after inflation, due to rapid
variation of it's effective mass (with frequencies having contributions 
from different scales, i.e. the Hubble constant, inflaton mass, and 
zero-temperature gravitino mass). This process is similar to the 
production of particles during a preheating era. However, due to the
complexity of supersymmetric theory, it is far from straightforward to
make predictions. It is thus not too surprising, that claims 
(for $m_{\phi}^2\phi^2/2$ inflation) range
from limits on $T_{\rm RH}$ of $\sim 10^{12}\,$GeV~\cite{KKLP00},
$\sim 10^{5}\,$GeV~\cite{GRT99},
to statements~\cite{L00}
that a short, secondary (thermal) inflationary period is required.
It has also been argued~\cite{KKLP00}
that inflation with potential $\lambda\phi^4$
is likely ruled out altogether due to this gravitino constraint.
However, the above claims have recently been 
challenged~\cite{NPS01}. It seems not
clear, if in this process of
``preheating'' indeed the spin (1/2) component of the gravitino is excited
or rather, the supersymmetric partner of the inflaton particle. 

\vskip 0.1in
\noindent 
All the above illustrates that ultimately a
\lq\lq low energy scale\rq\rq ($H$ small) inflation, and/or low cosmic
reheating temperature may be required.  Alternatively, 
unwanted relics generated at the end of inflation may subsequently
be diluted by a secondary epoch of thermal inflation.

\subsection{Wimpzillas}

\begin{figure}[t]
\centering 
\leavevmode\epsfysize=7cm \epsfbox{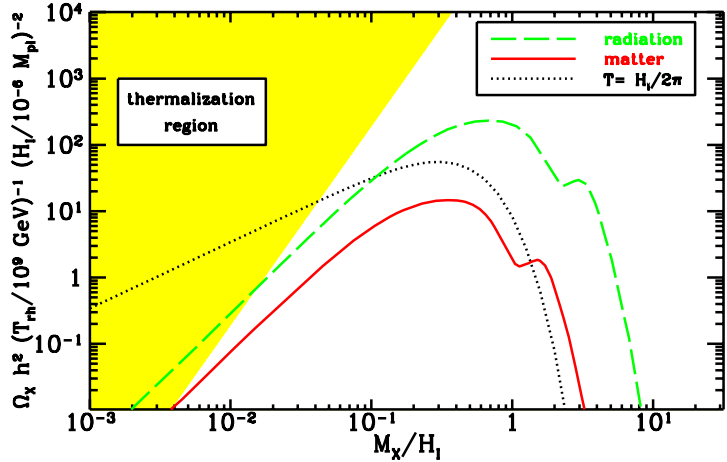}\\ 
\caption[scdm]{Fractional contribution to the present critical
density of a conformally coupled boson produced out of vacuum due
to the rapid expansion rate $H_I$ during the final stages of an 
inflationary epoch. The reader is referred to Ref.~\protect\cite{CKR99} for details.}
\label{fig5}
\end{figure}

Production of very massive, weakly interacting particles via
non-equilibrium processes in the early Universe, such as during preheating,
or by production out of the vacuum via gravitational phenomena,
as discussed in Sec. 6.1.2., may be of interest for the cosmological dark matter
problem~\cite{CKR99,KT98}. 
Such particles have been coined WIMPZILLAs~\cite{KCR98} 
and circumvent the
well-known unitarity bound on the mass of a dark matter particle 
$M_{\rm dark} \,{}^<_{\sim}\, 200\,$TeV~\cite{GK90}, which may be
left over from a freeze-out from an initially equilibrium abundance
(of which an excellent candidate is the lightest supersymmetric particle).
Whereas production of the right abundance of WIMPZILLAs
during preheating may be subject to
fine-tuning, it may be seen from Eq.~(\ref{Omega}) (and the remarks below it)
that for fermions, or conformally coupled bosons, of mass 
$m_{\chi}\sim 10^{13}\,$GeV and with the canonical reheat temperature 
$T_{\rm RH}\sim 10^9\,$GeV (to avoid gravitino overproduction) a fractional
contribution to the present critical density of $\Omega_{\chi}^0\sim 1$
naturally results. This is illustrated in Fig.~\ref{fig5}.
If unstable, with life time $\sim 10^{20}\,$yr
such particles may also have implications for the
observed highest-energy cosmic rays~\cite{HECR}. If, indeed, WIMPZILLAS are
the dark matter, detection of them may turn out to be exceedingly difficult.

\section{Q-balls - dark matter and baryogenesis}

Supersymmetric theories generically contain many
complex scalar fields $\varphi$, symmetric under $U(1)$ transformations. 
The charges $Q$ associated with these symmetries
represent baryon-number and lepton number (and possibly other unknown
conserved quantum numbers). Many of these fields have either, flat 
potentials (e.g., typical in gauge mediated supersymmetry breaking at 
some low scale),
or potentials approaching $\to\varphi^2$ from below
(e.g., typical in gravity mediated supersymmetry breaking at 
some intermediate scale),
as $\varphi$ approaches infinity.
In the early Universe, such fields are expected to be displaced from the 
origin during inflation, and if sufficiently ``light'', are not damped
away during the inflationary process (cf. Sec. 6.1.1.).
The resulting baryon- (lepton-) number carrying condensate after inflation,
may decay, thereby generating cosmic baryon (lepton) number. 
This process, envisioned to occur
within the homogeneous condensate, is known as Affleck-Dine 
baryogenesis~\cite{AD85}. 
Nevertheless, it has been shown that the condensate may be 
physically unstable towards
fragmentation~\cite{KS98}. 
Numerical analysis confirms this phenomenum~\cite{KK00}. 
The fragmentation process leads to the 
formation of non-topological solitons~\cite{Qreview}
 - B-balls (carrying baryon number), 
or L-balls (carrying lepton number).
Depending on the details of supersymmetry breaking, B-balls may be 
stable since their mass may be smaller than that of a collection
of nucleons (with individual mass $m_N$ and with the same baryon number $B$), 
i.e. B-ball mass $M_B$ is given by 
$M_B \approx M_s B^{3/4} < m_NB$ for large $B$ contained
in the soliton. This typically occurs 
when $V(\varphi) \to M_s^4$ as $\varphi\to\infty$.
They may also be unstable, but long-lived, e.g. 
$M_B \approx M_s B > m_NB$, when $V(\varphi) \to M_s^2\varphi^2$.  
In contrast, L-balls are generally unstable, and are unlikely to be
a remnant of the early Universe, due to efficient emission of (almost)
massless neutrinos from the soliton.

The existence of Q-balls in the spectrum of supersymmetric theory
may have implications for the cosmological
baryogenesis and dark matter problems. Stable (at zero-temperature) B-balls
may generate the baryon asymmetry by partial (or complete) evaporation
in the early Universe. Moreover, when evaporation occurs partially only,
the remaining solitons may act as the dark matter~\cite{KS98,LS98}
\footnote{with possibly interesting interactions from a structure formation
point of view~\cite{KS01}}. They thus could provide
explanation for both, baryogenesis and dark matter, within the same scenario.
Similar may happen with unstable B-balls~\cite{EM}. 
When supersymmetric R-parity is
conserved, evaporation of unstable B-balls in the early Universe may
generate an in principle calculable ratio between cosmic baryon number
and lightest supersymmetric particle (LSP) number (e.g. neutralinos). If decay
occurs sufficiently late $T\siml 100\,$GeV, 
no further reprocessing of these abundances
result. Since the expected ratio LSP/B is of order unity, such scenarios
may accommodate a cosmic coincidence, in particular, that
$\Omega_{DM}$ and $\Omega_B$ are of the same order of magnitude.
This fact is usually not addressed in other scenarios of baryogenesis.

\section{Primordial black holes}

It has long been known that primordial black holes (PBHs) may form
from radiation in the early Universe when only moderate density 
fluctuations ($\delta\rho /\rho\approx 1$, where $\rho$ is density) on
the cosmological (and time-dependent) horizon scale exist~\cite{PBH}. 
(i.e. $R_H\approx 2ct$,
where $t$ is cosmic time and $c$ is speed of light)
\footnote{The reader is referred to Sec. 5.2 for another possibility
of PBH formation}. This provides
``observational'' support for a relatively homogeneous 
early Universe (especially after inflation), since if
PBHs would form early on, they may either easily dominate cosmic
energy density of the Universe at an too early stage, or be in conflict 
with observations due to observable effects of their evaporation. With the 
help of this argument,
the primordial density spectrum (as, for example, generated during inflation)
may be constrained from arguments pertaining to extremely small 
spatial scales~\cite{PBHconst}
(as compared to CMBR scales). The scales where constraints may be derived
from PBH formation may
even be further decreased when the possible evaporation of 
supersymmetric particles are considered~\cite{PBHsmall}
\footnote{This may not be the 
case, if, as recently claimed, a photosphere
forms during the PBH evaporation process~\cite{photo}.}.
On the other hand, PBHs may still be a viable
dark matter candidate. In particular, PBHs of approximate mass scale
$M_{BH}\sim 1\, M_{\odot}$ would form during the QCD epoch. Due to 
a soft equation of state because of the color confinement transition, 
PBHs form more easily during this epoch than during
other radiation dominated eras of the early Universe~\cite{J97}. This
could provide explanation for the as yet unexplained findings of the 
MACHO and EROS collaborations, that a fraction of the galactic Milky Way 
halo {\it may} be in form of compact $\sim M_{\odot}$ 
objects~\cite{micro}.
Finally, there has been progress in the computation of
PBH mass functions in the early Universe. In the early nineties discovered
critical phenomena in general relativity may, in fact, be exploited
to derive the masses of black holes~\cite{NJ98}, i.e.
$M_{BH} = k(\delta - \delta_c)^{0.35}$, given initial data of
overdensity $\delta\equiv \delta\rho /\rho$ and a calculated critical
overdensity $\delta_c$ for the threshold of the PBH formation process. 
Here $k$ is in principal calculable, typically
being of the order of the horizon mass during the PBH formation process.

\section{Big Bang nucleosynthesis: observational progress}

Big Bang nucleosynthesis (BBN) is one of the main pillars of the
hot Big Bang model. There exist several excellent (and recent) reviews
on the subject, either on the standard model~\cite{SBBNreview} 
employing minimal
assumptions (see below), or on the many possible non-standard variants of 
BBN~\cite{S96,NBBNreview}, where one (or several) of these assumptions is 
relaxed. The reader is principally referred to these reviews, 
or the original literature. In what follows
I will, nevertheless, summarize theory (in the standard model)
and recent observational progress briefly.

BBN leads to the production of the bulk of $^4$He, (probably) all D, 
and good fractions of $^3$He and $^7$Li observed in the Universe at present.
It is the earliest period of the early Universe where 
``some sort'' of precision statements may be made, by detailed comparison
between observationally inferred and theoretical predicted primordial 
light-element abundances. BBN occurs as a freeze-out process from
nuclear statistical equilibrium of elemental abundances, in particular,
between the epoch of weak freeze-out 
$T\approx 2-3\,$MeV (where interconversion between neutrons 
$n$ and protons $p$ is too slow to maintain equilibrium n/p-ratios) 
and a subsequent epoch at $T\approx 10 - 20\,{\rm keV}$,
when cosmic densities and temperature have dropped sufficiently for nuclear
reaction being essentially frozen out altogether. Making detailed
predictions for the resulting light-element abundances involves making
a number of assumptions about cosmic conditions, and neutrino properties.
In the standard model, these are uniform baryon density, the 
existence of three, light, stable, and nondegenerate 
(i.e. no significant asymmetry between neutrino- and anti-neutrino 
numbers) active neutrino species\footnote{neutrino oscillations may
also have impact on BBN yields, either when mixing occurs with a fourth
sterile neutrino, or when neutrino degeneracy is present},
a \lq\lq clean\rq\rq\ universe, i.e. no relic (decaying) particles, no
pockets of antimatter, no defects or evaporating black holes, as well as
the applicability of classical relativity and non-varying physical 
constants (e.g. fine structure constant 
$\alpha$) in the early Universe. Under these assumptions, there is but one
free remaining parameter, the cosmic baryon-to-photon ratio 
$\eta = n_B/n_{\gamma}$ directly related to $\Omega_B$,
which specifies the model. Note that the large number of assumptions in 
this minimal model exemplifies the constraining power of BBN. 
Nevertheless, full use 
of this may only be made when 
detailed calculation of non-standard BBN have been performed.

\begin{figure}[t]
\centering 
\leavevmode\epsfysize=10cm \epsfbox{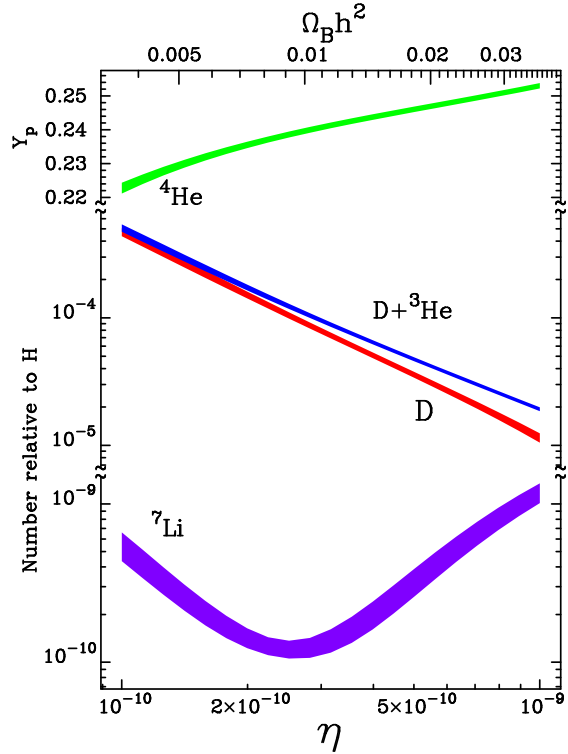} 
\caption[scdm]{Predicted light-element abundance
yields in a standard BBN model as a function of baryon-to-photon ratio.
Refer to Ref.~\protect\cite{BNT01} for details.}
\label{fig6}
\end{figure}

In Fig.~\ref{fig6} 
theoretical predictions for the synthesized abundances
in the standard model as a function of $\eta$ are shown. Here the width
of the bands indicate theoretical uncertainties ($95\%$ confidence)
in these predictions,
which, with the exceptance of the $^7$Li isotope, are negligible when
compared to uncertainties in their observationally inferred values. 
The field of BBN has recently been revolutionized by the ability of making
comparatively accurate abundance ratio determination of D/H within
absorption line systems on the line-of-sight towards distant, high-redshift
quasars. Currently the ``cleanest'' system (towards QSO 1937-1009)
shows abundance of D/H $= (3.3\pm 0.6)\times 10^{-5}$~\cite{BT98a}. 
This, together with two further 
claimed accurate determinations of $4.0\pm 1.4\times 10^{-5}$~\cite{BT98b},
and $2.5\pm 0.5\times 10^{-5}$~\cite{OMetal}, 
has been taken to derive the value 
of $3.0\pm 0.2\times 10^{-5}$ for the primordial D/H ratio.
Whereas initial determinations have been attempted mostly within
lower column density ($N_H\sim 10^{17}\,{\rm cm}^{-2}$) Lyman-limit
systems, inferred D/H ratios within higher column density 
($N_H \sim 3\times 10^{20}\,{\rm cm}^{-2}$) damped Lyman-$\alpha$
systems have recently also
been put forward. These include
D/H $= (1.65\pm 0.35)\times 10^{-5}$~\cite{PB01} and 
D/H $= (3.75\pm 0.25)\times 10^{-5}$~\cite{LDOM01}.  
Inferred D/H ratios 
\footnote{assumed to be close to primordial due to 
mostly theoretical constraints on the
amount of deuterium destruction in these clouds},
when plotted against column density, are consistent with an unexpected
trend of decreasing D/H-ratios with increasing $N_H$~\cite{BNT01a}. This
illustrates that the subject of determinations and interpretations
of D/H-ratios in Lyman-$\alpha$ absorbers has not completely settled
as yet.

Comparatively accurate determinations of primordial $^4$He and $^7$Li
abundances may also be made. From a large sample of extragalactic
HII-regions and compact blue galaxies a value of 
$Y_p = 0.244\pm 0.002$ for the primordial helium mass fraction has been
derived by one group~\cite{YT98}. This may be compared to the result of 
$Y_p = 0.234\pm 0.002$ by another group, based on a similar 
sample~\cite{OSS97}, as well as the result inferred by a careful 
analysis of one single HII region, $Y_p = 0.2345\pm 0.0026$ by
two of the pioneers of the field of abundance determinations in
gaseous nebulae~\cite{PPR00}. The spread in results, together with
theoretical estimates of the possible systematic uncertainties in the
procedure of observational primordial $^4$He 
determinations~\cite{4Hesys}, illustrate that the quoted error bars of all
groups represent most certainly a significant underestimate of the
true intrinsic errors of the method. Note that, within a standard
BBN model, a large $Y_p$ abundance (as given above) is consistent with the 
inferred  ``average'' D/H $\approx 3\times 10^{-5}$ ratio, whereas a small one
is not. This fact sometimes seduces cosmologists 
to ``prefer'' the large $Y_p\approx 0.244$ value. Concerning the $^7$Li
isotope, abundance determinations may be made in halo stars belonging
to the Spite plateau, known to have constant $^7$Li/H 
ratios in different stars, with a value believed to be intimately connected
to the primordial $^7$Li/H abundance. Two recent results are given by
{$^7{\rm Li/H} = 1.73\pm 0.25\times 10^{-10}$}~\cite{BM97}
and {$^7{\rm Li/H} = 1.23^{+0.68}_{-0.32}\times 10^{-10}$}~\cite{Retal00}.
Here the latter group claims to observe, for the first time, a contribution
of $^7$Li in the atmospheres of these stars due to synthesis of 
$^7$Li via galactic cosmic ray induced spallation processes. 
Nevertheless, both quoted values are inconsistent within the standard model
with a D/H-ratio of $3\times 10^{-5}$ (which predicts $^7$Li/H
$\approx 3.8\times 10^{-10}$ cf. Fig.~\ref{fig6}). This fact holds true even
when theoretical uncertainties in the predicted $^7$Li/H ratio are accounted
for. It seems thus that stellar $^7$Li depletion,
reducing the primordial $^7$Li
abundance in stars on the Spite plateau by a factor $\sim 2$, is required.
(Alternatively, it may be that most D/H determinations represent an
underestimate of the true primordial D/H abundance). 

In conclusion, though there may exist certain mismatches
between observationally inferred and theoretically predicted primordial 
abundances within the standard BBN, these are probably 
accounted for by still appreciable systematic uncertainties
in individual primordial abundance determinations. There exists therefore 
currently no strong motivation to move to non-standard models. Given a value
of D/H $= 3.0\pm 0.2\times 10^{-5}$, one may derive for the
fractional contribution of baryons to the critical density    
$\Omega_bh^2 = 0.02\pm 0.002$~\cite{BNT01,OMetal},
where $h$ is the Hubble parameter in units
of $100\, {\rm km}\, {\rm s}^{-1}{\rm Mpc}^{-1}$. 
This value is probably also consistent with
recent determinations of $\Omega_b$ from the CMBR~\cite{CMBnew}. On the other
hand, earlier quoted values of $\Omega_bh^2$ from observations of
CMBR anisotropies~\cite{CMBold} 
as large as $0.03$ may simply not be accommodated 
within a standard BBN model. 

\section{Conclusion}

In conclusion, let me just make a short remark. Early Universe cosmology
in the eighties discovered the virtues of a (Hubble) damping term in the
cosmological equation of motion for bosonic fields. This discovery led to
the invention of inflation. In the nineties the virtues (and problems) of
terms leading to instabilities in the equation of motion for bosonic fields
in the Universe were discovered. This revelation led to the possibility of
cosmological preheating and formation of Q-balls. Inspection of the
remaining terms in the equations of motion for a bosonic
field in the Universe may give, the educated reader, an outlook on possible
developments in the first decade of this millennium.

\ack
The author thanks the organizers of the conference for invitation 
and financial support, and the opportunity to visit South Africa. 
I am also grateful to my girlfriend Nanci,
who ``forced'' me to be a responsible scientist and write 
this article.

\end{document}